\documentstyle[12pt, fleqn,epsf,epsfig]{article}

\title{Functional Dynamics II:\\Syntactic structure}
\author{
Naoto Kataoka and Kunihiko Kaneko\\
  {\small \sl Department of Pure and Applied Sciences}\\
  {\small \sl University of Tokyo, Komaba, Meguro-ku, Tokyo 153, JAPAN}\\
}
\date{}

\begin{document}
\maketitle

\abstract

Functional dynamics, introduced in a previous paper, is
analyzed, focusing on the formation of a hierarchical rule to determine the
dynamics of the functional value.  To study the periodic (or non-fixed)
solution, the functional dynamics is separated into fixed  
and non-fixed parts.
It is shown that the fixed parts generate a 1-dimensional map 
by which the dynamics of 
the functional values of some other parts are determined.
Piecewise-linear maps with multiple branches are generally created,
while an arbitrary one-dimensional map can be embedded into this functional dynamics if
the initial function coincides with the identity function over a finite interval.
Next, the dynamics determined by the one-dimensional map 
can again generate a `meta-map', which
determines the dynamics of the generated map.
This hierarchy of meta-rules can continue recursively.
It is also shown that the dynamics
can produce `meta-chaos' with an orbital instability that is stronger
than exponential.
The relevance of the generated hierarchy to
biological and language systems is discussed, in relation
with the formation of syntax of a network.

\section{Introduction}

In a previous paper \cite{I} (to be referred as I),
we introduced functional dynamics to investigate
the articulation process carried out on an initially inarticulate network.
In the functional dynamics, objects and rules are not separated 
in the beginning, and we study how objects and rules appear from an
inarticulate network through iterations of functions.
The functional dynamics provides a simple universal model for this appearance.

In the general introduction of the paper I, we discuss five requisites for a biological system,
while two of them are explicitly studied. These are the following:
\begin{itemize}
\item
Inseparability of rule and variable ($f\circ f$).
\item
The articulation process from a continuous world.
\end{itemize}
In I, we investigate functional dynamics defined by
\begin{equation}
f_{n+1}(x)= (1-\epsilon) f_n(x)+\epsilon f_n \circ f_n(x)
\end{equation}
The evolution of the function $f_n(x)$ has been studied with $n$ representing the iteration step and
$\epsilon$ as a parameter. 
We have shown that an articulation process is
generated in this 1-dimensional functional dynamics.
As $f_n(x)$ evolves, first, type-I fixed points satisfying $f(x^I)=x^I$ are formed, and then
type-II fixed points satisfying $f(x^{II})=x^I$ are formed \cite{I}.
The articulation process is studied as
a classification process of how $f_n(x)$ converges to distinct intervals
consisting entirely of type-II fixed points.
For a given value $a = f_n(x)$, the inverse set $I_n = f^{-1}_n(a)$ is given
as an articulated class.
This means that the filter articulates the continuous world $x$
into some segments according to the value $f_n(x)$.
For such sets, $I_{n+1} \supseteq I_n$ holds, and
the dynamics of this system is determined completely by a set of relations among these
intervals as $n \rightarrow \infty$.  This reduction of the degrees of freedom out of a continuous world
is the articulation process.
This articulation is most clearly seen in the relation between
type-I and type-II fixed points.  Intervals of type-II
points corresponding to the same type-I fixed points are generated
from an initial continuous function.

As an articulation process, a structure independent of $n$ is formed by the fixed points, while
the functional values of some points change periodically
in time, taking the values of different type-II fixed points successively
(i.e., being mapped to the rigid structure constituted by fixed points).
This periodic function provides an example of how objects and rules depend
on each other, based on a rigid structure unchanged under iteration.

In the present paper, we focus on a dynamical aspect of these
functional dynamics, to study how rules for dynamic change emerge 
through iteration of the functional mapping.  With this approach, we investigate
the third and forth requisites mentioned in the general introduction of I:
\begin{itemize}
\item
Formation of a rule to change the relation among objects.
\item
Formation of hierarchical rules.
\end{itemize}

From our viewpoint, objects and rules emerge from the same level in which
code and encoding circulate dynamically.
For example, in natural language, there is a set of words and rules that 
forms sentences.
When we assume in the beginning that the objects and rules are already separated,
the theory of language has to be based on 
formal languages \cite{Chom}\cite{Hopc}, since the structure of the language has to be studied
without resorting to the objects.  With the separation of words and rules, 
one neglects the fact that the rules have to be described by using words, while 
the meaning of a word has to be described by a sentence.
This implies that a theory starting from a hierarchy in which words and objects are separated 
is not sufficient as a mathematical framework for natural language.  
Natural language is described
as an assembly of objects, rules, meta-rules, meta-meta-rules, and so forth,
while this hierarchy is not given in advance.
Indeed, this process of emergence of hierarchical structure from pre-structured ``something"
is a common characteristic of biological systems, as is seen, for example, in the hierarchy of
cell, tissue, organism, and so forth, starting from
an assembly of chemical reactions.  A mathematical formulation is required
to study the hierarchy of the successive formation of rules 
at successively higher levels \cite{Varela} \cite{Rosen}.

By taking the same viewpoint as that in I, 
we study a network of language/objects as a functional form.
With the evolution of the  functional dynamics, some structure
is constructed step by step.  In particular, we study how
a hierarchy of rules and meta-rules emerges through the
iteration.  Here, a structure is formed first by the configuration of
fixed points, while a
rule for the dynamic change of the structure is organized according to an `orbit' of the functional
dynamical system, and then a meta-rule is formed governing the
dynamic structure generated by the orbit.

This implies that once we get a rigid (fixed-point) structure, unchanged under repeated mapping,
as the elementary part of the dynamical network,
hierarchical structure appears under some restrictions.
Here, the rigid structure and hierarchical structure correspond to 
the articulated objects and the operator acting on the objects, respectively.
This separation of objects and rules emerges, 
since we extract the rigid structure out of the functional dynamics.

The present paper is organized as follows.
In Sec.2, we explain the basic properties of the functional dynamics again 
to facilitate its representation by the introduction of a `generated map'.
With the generated map, it is shown in Sec.3.1, that  a 1-dimensional map is
embedded in the functional dynamics. This map works as a rule governing the
change of the functional values over some intervals.
In Secs.3.1 and 3.2, piecewise linear maps
called the Nagumo-Sato map and the `multi-branch Nagumo-Sato map' are naturally embedded 
in the functional dynamics.
In Sec.3.3  a larger class of 1-dimensional maps is embedded 
into our functional map. This implies that chaotic functional dynamics is possible in our system.
By choosing a suitable initial condition, it is shown in Sec.4 that the functional
dynamics can form a hierarchical structure.  A meta-rule for the change
of the functional values is formed which changes according to the (chaotic) dynamics
generated by the 1-dimensional map.
The maps can be nested recursively and generate higher level meta-maps successively.
In Sec.5, we discuss syntactic structure derived from this functional dynamics,
and the relevance of our results to the 3rd and 4th requisites of biological systems,
mentioned above.

\section{Model}

The functional map Eq. (1) has the form

\begin{equation}
f_{n+1}(x) = F(f_n(x), f_n \circ f_n(x)).
\end{equation}
Here we study some characteristics of this functional equation with a 1-dimensional $f_n(x)$.
In connection with our motivation for biological systems and language structure,
the function \(f_n(x)\) is considered to represent an abstract relation network, while
\(f_n \circ f_n(x)\) provides  a self-referential term.
Since we are interested in modeling the
situation in which code and encoding are not separated,
\(f_n(x)\) represents a projection from a set into itself.

First, we discuss two characteristic properties of this equation:
\begin{itemize}
\item
If $x'$ and $x''$ have the same value $f_n(x') = f_n(x'')$ at $n$,
the subsequent evolutions of $f_m(x')$ and $f_m(x'')$ (with $m > n$) are identical,
because dynamics are determined completely by the function $f_n(x)$. 
\item
The equation (1) can be split as
\begin{equation}
\left\{
\begin{array}{lcl}
f_{n+1}(x) & = & g_n (f_n(x))\\
g_n(x)     & = & F(x, f_n(x)).
\end{array}
\right.
\end{equation}
\end{itemize}
The first property above implies the ability of articulation of this system.
Once $f_n(x)$ identifies $x'$ and $x''$ as the same thing, 
the two points evolve in the same way.
The second property provides a novel viewpoint to study this model.
With this separation, one can say that a point $f_n(x')$ 
evolves under \(g_n(x)\), which is generated from \(f_n(x)\) itself.
This is a characteristic property of this functional equation.
Given $f_n(x)$, a map $g_n(x)$ is determined. 
The $f_n(x)$ evolves to $f_{n+1}(x)$ under the map and $g_{n+1}(x)$ is determined.
In this paper, the term `function' is used to represent \(f_n(x)\),
while a `generated map' is used in reference to \(g_n(x)\).

If $f(x')$ is a fixed point at $x'$
(which does not mean a fixed function as a whole),
the generated map of  $f(x')$ is a fixed generated map at the point $x'$.
To study the functional dynamics,
we first study a fixed generated map $g(x)$ and see how other points $x''$
evolve under this generated map (Sec.3).
Second, we study the case in which $g_n(x)$ itself changes in time,
by taking a suitable initial configuration of $f_0(x)$.  There,
a hierarchical structure (meta-map) is considered (Sec.4).\\

For simplicity, we impose one more restriction on (2),
following I.
We assume that, if the relation $f(x')$ is fixed, it satisfies 
the relation $f(x') = f \circ f(x')$.
This condition implies that the change of $f(x')$ vanishes when the self-reference of a function
agrees with the function itself.
The simplest model of this type is (1), obtained
by choosing the form

\begin{equation}
F(x, y) = (1-\epsilon)x + \epsilon y,
\end{equation}
with $0 < \epsilon < 1$.
(The case with a general form for $F(x, y$) 
is briefly discussed in Appendix A and 
will be discussed in a future paper.)

For the type of model we study, the dynamics relax toward the self-consistent relation
$f(x') = f \circ f(x')$.
%Now, the equation (1) has the form;
%\begin{equation}
%f_{n+1}(x) = (1 - \epsilon)f_n(x) + \epsilon f_n \circ f_n(x),
%\end{equation}
%which is adopted in the previous paper I also.
For the remainder part of this paper,  we focus on the functional dynamics (1).
In this case,
the generated map is given by
$g_n(x) = (1 - \epsilon)x + \epsilon f_n(x)$.

Now we obtain two useful properties:
\begin{itemize}
\item A value $f_n(x')$ which satisfies the condition $f_n(x') = f_n \circ f_n(x')$ is a fixed point.
\item There is a transformation $T$ which satisfies the condition $F \circ T = T \circ F$.
(The explicit form of $T$ is discussed below.)
\end{itemize}

Here, all the points $x^I$ which satisfy $f(x^I) = x^I$ are 
fixed points of the functional equation(1). 
Since all points $x$ with an identical value $f(x)$ evolve identically,  
all the points $x^{II}$ that satisfy $f(x^{II}) = x^I$ are again fixed points. 
For convenience, we have classified (see I) these fixed points as 
\begin{itemize}
\item $x^I$ is a type-I fixed point with \(f_n(x^{I}) = x^{I}\) .
\item $x^{II}$ is a type-II fixed point with \(f_n(x^{II}) = x^{I} \neq x^{II}\).
\end{itemize}
A type-I fixed point is a point 
at which \(f_n(x)\) intersects the identity function. 
%while a type-II fixed point is a point that has the same value 
%\(f_n(x)\) as some type-I fixed point.
This `type' is extended to arbitrary type-$N$.
We define a type-$N$ point as a point which satisfies the condition 
$f(x^N) = x^{N-1}$, after the transient in the
functional dynamics has died away.
Here $x^N$ represents a type-$N$ point.
Although type-I and II points are fixed points, 
type-$N$ ($N > 2$) points cannot be fixed points.
In fact, if $x$ is a fixed point and $y=f(x)$, 
the fixed point condition is written $y=f(y)$, which 
means $y$ is a type-I fixed point and $x$ is a type-I or II fixed point.

In I, we introduced the concept of a `self-contained section' (SCS),
which is defined as a connected interval $I$ such that 
$f(I) \subset I$, while no connected interval $J \subset I$ satisfies $f(J) \subset J$,
and $f(I+\delta) \subset I+\delta$ for arbitrary small $\delta$, either.
Here, we extend this definition to introduce the `closed section' (CS) and 
`closed generated map' (CGM).  A CS is defined as a set $I$ such that 
$f(I) \subset I$ (where $I$ is not necessarily connected), while
a CGM is defined as a set $J$ such that 
$g(J) \subset J$ (where $J$ is not necessarily connected).
In Eq.(1), if $f_0(I) \subset I$,
then $f_n(x) \in I$ for all $n$, and $I$ is a CS.\\

Now let us return to the transformation $T$. 
The dynamics of this functional equation is invariant by the transformation $T$:
\begin{equation}
\left\{
\begin{array}{lcl}
 ax + b      & \rightarrow & X\\
af_n(x) + b  & \rightarrow &F_n(X).
\end{array}
\right.
\end{equation}
In fact, Eq.(1) assumes the form of an operation of taking a weighted average of  
\(f_n(x^{\prime})\) and \(f_n \circ f_n (x^{\prime})\).
Thus, the functional dynamics are invariant under a
scaling transformation in which $x$ and \(y=f(x)\) are multiplied by the same factor
and shifted by the same value.
This invariance means that $T$ and $F$ commute ($T\circ F = F\circ T$).
Under this transformation, a connected CS (CGM) $(x_1, x_2)$ 
is shifted to $(ax_1 +b, ax_2+b)$ giving
a new CS (CGM).
The above invariance will be used to embed a 1-dimensional map into this functional dynamics in Sec.3
and to construct a meta-map in Sec.4.

\section{1-Dimensional Map in Functional Dynamics}

\subsection{General Properties}

In I, 
we found that the function $f_n(x)$ does not converge to a 
fixed function as $n \rightarrow \infty$ for some initial functions.  
For example, for the initial
function $f_0(x) = rx(1-x)$, $f_n(x)$ does not converge to a fixed 
function for some range of parameter $r$ referred to as the R (random) phase
in I, where the number of discontinuous points of $f_n(x)$ and the length 
of $f_n(x)$ increases in proportion to $M$, the number of mesh points 
adopted for the numerical calculation.

\begin{figure}
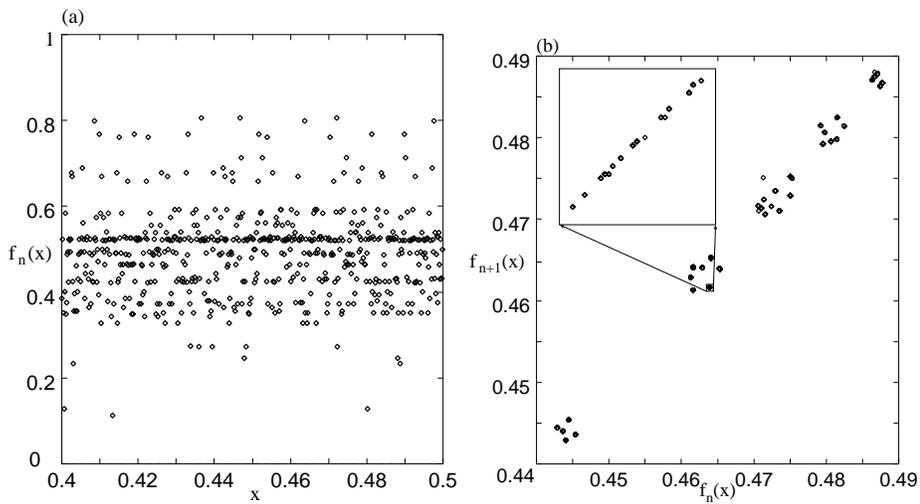

%\begin{center}
\noindent
\epsfig{file=s.eps,height=6cm,width=6cm}
\epsfig{file=ret.p.eps,height=6cm,width=6cm}
\caption{(a) The $f_{50000}(x)$ $(0.4 < x < 0.5)$ for random initial $f_0(x)$ with $M = 6000$ and $\epsilon = 0.02$.
It consists of type-I fixed points, type-II fixed intervals and some periodic points. 
(b) A part of the return map of (a) for all $(f_n(x), f_{n+1}(x))$ $(n = 50000-50050)$. 
The return map consists of some points and 
lines that have slope $(1 - \epsilon)$ (in the inset).
}
\label{retR}
%\end{center}
\end{figure}

Recalling that the $f_n(x)$ for large $n$ looks almost random in the R phase,
we have also computed $f_n(x)$ from Eq.(1) using
random initial conditions, as an extreme case.
For such initial conditions, we divide the interval 
$[0, 1]$ into $M$ intervals as
($[i/(M-1), (i+1)/(M-1)]$) and
choose the value of $f_n(i) \in [0, 1]$ randomly.
An example of $f_{\infty}(x)$ for an interval $x$ and the return map 
$(f_{n}(x),  f_{n+1}(x))$ for the interval are displayed in Fig.\ref{retR}.
Here, $f_n(x)$ mainly consists of many flat intervals with the same value,
while for some points $x$, $f_n(x)$ changes periodically in time.
As plotted by the return map, it is found that
the periodic dynamics obey a certain rule.  
As shown in the inset of Fig.\ref{retR}(b), a clear piecewise-linear
structure is visible in the return map.
In this section, we study how this type of return maps
is generated (Sec.3.2) and investigate the class of maps that can
appear with these functional dynamics (Sec.3.3).

A fixed generated map can be constructed from type-I and type-II fixed points.
This fixed generated map can act as a 1-dimensional map $g(x)$.
To extract the temporal change of other $x$ points,
it is useful to think of the interval $I$ as the union of three parts: 
$I = I_{map}\sqcup I_{driven}\sqcup I_{rest}$.\footnote{
As will be shown, the evolution of $f_n(x')$ for $x' \in I_{driven}$ is 
determined by $\{g(x)\}$ with $x \in I_{map}$.
In this sense, we call this interval `driven' by the interval $I_{map}$.}
Here $f(x)$ with $x\in I_{map}$ generates a fixed generated map $g(x)$ for 
$f_n(x)$ with $x \in I_{driven}$ 
and $f_{n+1}(x) = g(f_n(x))$ with $x\in I_{driven}$. 
Since $f(I_{map})$ is a fixed function, 
$f(I_{driven}) \subset I_{map}$ and 
$f(I_{rest}) \subset I_{driven}\sqcup I_{rest}$.

First, we assume there exists only one type-I fixed point $x^I$ (\(f(x^I) = x^I\)) 
and one type-II fixed point  $x^{II}$ and that $f(x^{II}) = x^I$.
The generated map at $x = x^{II}$ is $g(x^{II}) = (1 - \epsilon)x^{II} + \epsilon x^I$,
which can be rewritten as $g(x) = (1 - \epsilon)(x - x^I) + x^I$ by inserting $x = x^{II}$.

Next, we assume that there is an interval $I_{map}$ consisting entirely of a single type-I fixed
point and corresponding type-II fixed points,
which satisfy $f(x'') = x^I$ for $x'' \in  I_{map}$.  
Assuming the existence of such an interval,
the generated map is given by $g(x) = (1 - \epsilon)(x - x^I) + x^I$
for this interval $(x \in I_{map})$.
The map $g(x)$ is a line with a slope $1 - \epsilon$ that
intersects the type-I fixed point.
This line that is used as generated map 
is determined by the configuration of type-II fixed points.
If the interval $I_{map}$ is continuous, for $x \in I_{driven}$
all $f_n(x)$ evolve to $f_{\infty}(x) = x^I$.

\begin{figure}
%\begin{center}
\noindent
\epsfig{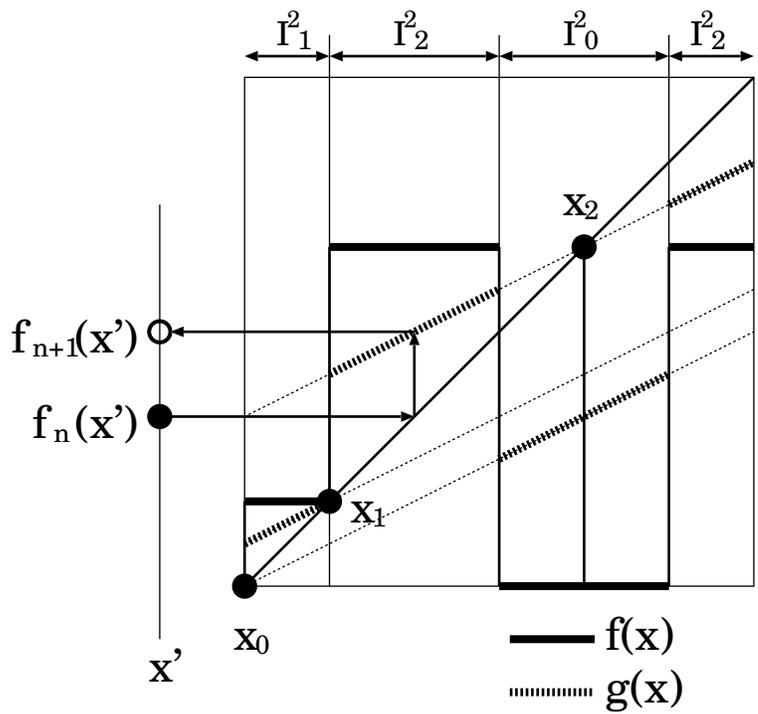}

\caption{An example of three type-I fixed points ($x_0 < x_1 < x_2$) and $f(x)$ for $x \in I$.
The map $g(x)$ is represented by the dotted line. $g(x)' = (1-\epsilon)$ and is constant.
the dynamics of another point $f_n(x') \in I$ is determined by $g(x)$.
}
\label{fig:multi}
%\end{center}
\end{figure}

Now consider the more general case in which
an interval $I_{map}$ consisting of several
type-I fixed points and several subintervals of type-II points that are mapped to 
one of the type-I fixed points.
In this case, the generated map is determined by the arrangement of type-I and type-II fixed points. 
This map is a piecewise linear function with slope $1 - \epsilon$, which intersect the type-I fixed points 
(see Fig.\ref{fig:multi}).
Here, we consider the following two cases for the configuration
of type-I fixed points:
\begin{itemize} 
\item There exist countable number of type-I fixed points (Sec.3.2).
\item There exist countable number of type-I fixed intervals (Sec.3.3).
\end{itemize}
A type-I fixed interval is a connected interval $[a, b]$ on which $f(x) = x$ for all $x \in [a, b]$.
A type-I fixed point is the limiting case of a type-I fixed interval (i.e., that in which $a = b$).

Let us denote the ordered set of type-I fixed points by $I^1 \equiv \{x_0, x_1, \ldots, x_{n-1}\}$
where $i < j$ implies $x_i < x_j$.
The ordered set of type-I fixed intervals are denoted in the same way as the fixed points, 
as $I^1 \equiv \{I^1_0, I^1_1, \ldots, I^1_{n-1}\}$,
where $i < j$ implies $\max I^1_i < \min I^1_j$.

Depending on the configuration of type-I and type-II fixed points,
a one-dimensional map is generated, which acts as a fixed CGM
for a point $f_n(x')$ for $x' \in I_{driven}$ 
In the next subsection, we study the case with isolated type-I
fixed points, while in Sec.3.3 we discuss the case with type-I fixed intervals.

\subsection{Case with Countable Type-I Fixed Points}

In this subsection, we consider the case with a
finite or countably infinite number of type-I fixed points.
As shown in I, the function $f_n(x)$ often tends to approach 
a piecewise constant function, 
consisting of a discrete set of type-I fixed points and several intervals 
of type-II fixed points at which $f(x)$ assumes the same value.
Thus the existence of such type-I fixed points and type-II intervals 
is common in our model (\cite{I} and Fig.\ref{retR}(a)).

Corresponding to type-I fixed points $x_i$,
we define sets of type-II fixed points $I^2_i$ to be those satisfying $f(I^2_i) = x_i$, where
$I^2_i$ is not necessarily connected and consists of several intervals in general.
Since $f(x)$ is a single-valued function, there is no intersection among $I^2_i$ and $I^1$.
The union of type-II fixed intervals ($\cup_i I^2_i$) is denoted as $I^2$.
Now, the interval $I$ is the union of $I^1, I^2, I_{driven}$ and $I_{rest}$
($I = I^1\sqcup I^2_0\sqcup I^2_1\sqcup \cdots \sqcup I^2_{n-1}\sqcup I_{driven} \sqcup I_{rest}$).
Following the argument in the last subsection,
the generated map in the interval $I_{map}$
has the form

\begin{equation}
g(x) = \left\{
\begin{array}{cccclcl}
g[0](x)   & =  &(1 - \epsilon)(x - x_0) + x_0            & \mbox{for }& x \in I^2_0,     & \mbox{where } &f(I^2_0) = x_0\\
g[1](x)   & =  &(1 - \epsilon)(x - x_1) + x_1            & \mbox{for }& x \in I^2_1,     & \mbox{where } &f(I^2_1) = x_1\\
\vdots    &    &         \vdots                          &           &  \vdots          &              &\vdots\\
g[n-1](x) & =  &(1 - \epsilon)(x - x_n) + x_{n-1}        & \mbox{for }& x \in I^2_{n-1}, & \mbox{where } &f(I^2_{n-1}) = x_{n-1}
\end{array}
\right.
\end{equation}
where $[i]$ denotes a line corresponding to the type-I fixed point $x_i$.
Each $g[i]$ is referred to as an `$i$-branch'.

As discussed above, this generated map acts as the evolution rule for
points $x'$ that are mapped to one of the type-II fixed points
[i.e., $f_n(x') \in I^2_i$, or, in other words,
$f_{n+1}(x') = g(f_n(x')) = (1 - \epsilon)f_n(x') + \epsilon x^I$
(see Fig.\ref{fig:multi})].

\begin{figure}
%\begin{center}
\noindent
\epsfig{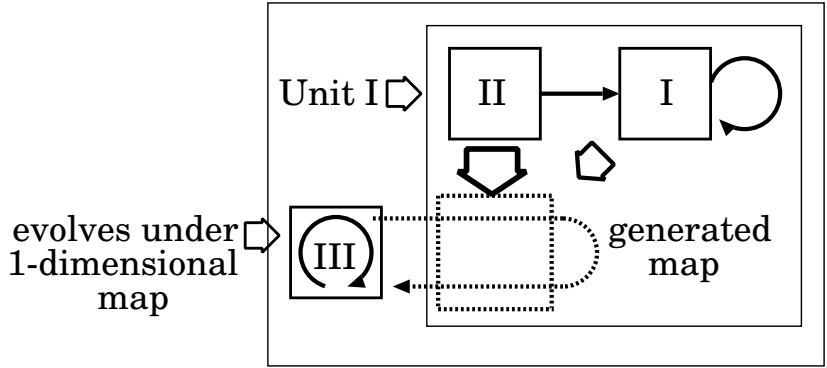}

\caption{The schema of the generated map.
The configuration of type-I and type-II fixed points generates the map, while the evolution of the  
type-III points are determined by the generated map.
}
\label{fig:sc1}
%\end{center}
\end{figure}

The combination of some type-I fixed points and an set of 
type-II fixed intervals satisfying certain conditions can give a CGM. 
Here we assume there exist $n$ type-I fixed points ($x_0 < x_1 < \cdots < x_{n-1}$),
and the points in the interval $(x_0, x_{n-1})$ are 
assumed to be mapped to one of the type-I fixed points.
Then, according to (6), for $i = 0, 1, \ldots, n-1$, $g[i](x') \in (x_0, x_{n-1})$ for all $x' \in I^2_i$.
As a total, $g(I^2) \subset I^2$, and the set on which $g(x)$ is defined is a CGM.
Thus, for $x \in I_{driven}$, the evolution of 
$f_n(x')$ $\in I^2_i$ is determined by the CGM.
This $f_n(x')$ is included in
a type-II fixed interval, and thus $f_n(x')$ can be called type-III.
The domain of $f_n(x') \in I^2_i$ is denoted as $I^3$ ($= I_{driven}$).
Now, the interval $I$ can be written as $I^1\sqcup I^2\sqcup I^3\sqcup I_{rest}$.
Here we call this type of configuration that generates a closed 1-dimensional 
map as `unit-I'.
The situation is drawn schematically in Fig.\ref{fig:sc1}.

For example, we assume that there are two type-I fixed points, 
\(x_0\) and \(x_1\) ($I^1 = \{x_0, x_1\}$).
We divide the interval $I = [x_0, x_1]$ into $I^1, I^2_0, I^2_1, I^3$ 
and $I_{rest}$.
Since $f(x)$ has only two values, $x_0$ and $x_1$, on $I^2$,
$g(x)$ has a `0-branch' and a `1-branch'.
The map \(g(x)\) has the same slope, (\(1-\epsilon\)), at each point 
\(x \in I^2\).  This class of map includes
the Nagumo-Sato map \cite{NS}.

\begin{figure}
%\begin{center}
\noindent
\epsfig{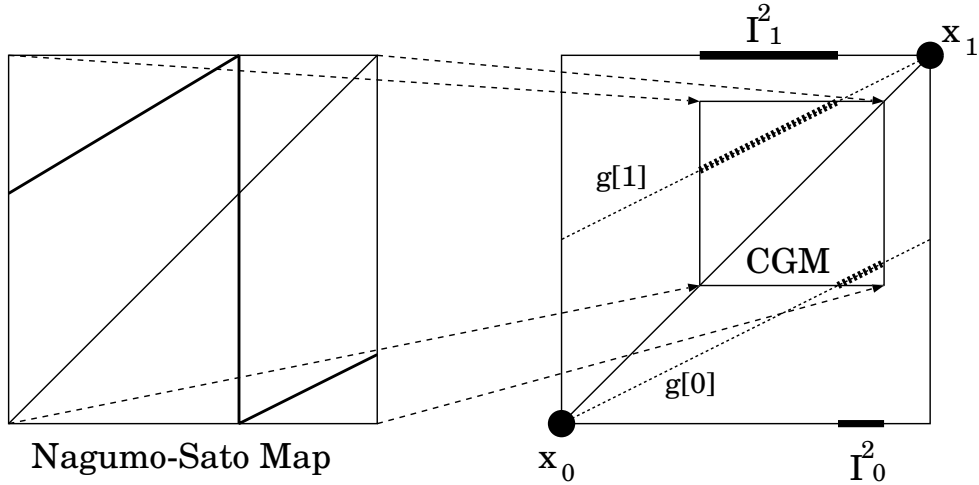}

\caption{An embedding of the Nagumo-Sato Map.
A transformation multiplying $\epsilon$ and moving $(1-w)$ along the $x$ and $y=f_n(x)$ directions
embeds a map into $g(x)$ for $\epsilon = 1-k$.
}
\label{fig:sato}
%\end{center}
\end{figure}

The Nagumo-Sato map is given by the equation
\begin{equation}
x_{n+1} = kx_n + w (\bmod 1),
\end{equation}
with \(0 < k < 1\) and \(0 < w < 1\).
This map has two branches with the same slope \(k\) for the interval
\([0, (1-w)/k]\) (the first branch) and 
\([(1-w)/k, 1]\) (the second branch)
(see Fig.\ref{fig:sato}).
To have these two branches, we need two intervals of type-II fixed points.
With the aid of transformation (5),
the domain of \(f_n(x)\) is restricted
to [0, 1], where two type-I fixed points are situated at 0 and 1,
without loss of generality.
Our purpose here is to show that certain
intervals $I^2_0$ and $I^2_1$ can generate a map of the form (7).

With the transformation (5), 
the slope $k = (1 - \epsilon)$ is conserved.
Transforming (7) by multiplying by $\epsilon$ and shifting by $1 - w$  
along $x$ and $y=f_n(x)$-directions,
we can embed the Nagumo-Sato map (of the interval size $\epsilon$) into $g(x)$.
The required condition is $I^2_0 = [(1-w)/(1-\epsilon), 1- \epsilon + w]$ and 
$I^2_1 = [1 - w, (1-w)/(1-\epsilon)]$ (See Fig.\ref{fig:sato}).
This map $g(x) \in [1 - w, 1- w + \epsilon]$ becomes a CGM. 
Note that this situation can generally arise without 
choosing a very special initial function.  This is why the functional  
dynamics from arbitrary initial conditions often lead to a periodic
cycle governed by the Nagumo-Sato map, as in  Fig.1.

\begin{figure}
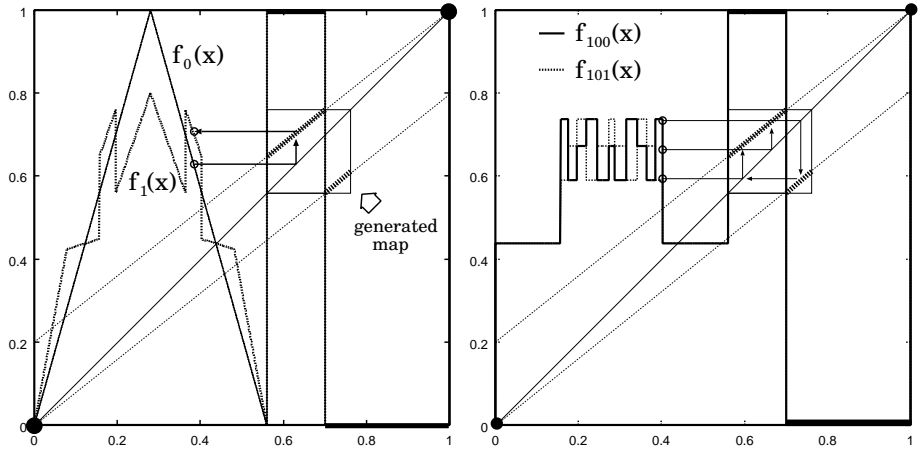

%\begin{center}
\noindent
\epsfig{file=ns.0-1.eps,height=6cm,width=6cm}
\epsfig{file=ns.101-102.eps,height=6cm,width=6cm}

\caption{Time evolution of \(f_n(x)\).
The function \(f_n(x)\) is divided into two parts, 
the map area and the rest.
Here, fixed intervals produce the Nagumo-Sato map, and 
the dynamics of \(f_n(x)\) which is mapped to the square is determined by the map.
Here \(a = 0.7\),\(\epsilon = 0.2\). 
At the square, the generated map $g(x)$ has the same properties as the map \(x_{n+1} = 0.8x_n + 0.44\).
(a) \(f_0(x)\) and \(f_1(x)\) are plotted.
$f_0(x)$ is chosen as described in the text.
The dynamics of the function mapped to the square region is determined by the generated map,
while the remaining part converges to a type-II fixed point.
(b) \(f_{100}(x)\) and \(f_{101}(x)\) are plotted.
All points converge to fixed or periodic points.
The periodic points are determined by the Nagumo-Sato map.
Each point is period-3, and as a whole \(f_n(x)\) is a period-3 function.
}
\label{fig:sato3}
%\end{center}
\end{figure}

An example of our
simulations is displayed for \(\epsilon = 0.2\) and \(w = 0.44\) 
in Fig.\ref{fig:sato3}, where
the discontinuous point ($a = (1-w)/(1-\epsilon)$) of the Nagumo-Sato map 
is located at 0.7.  In the simulation,
the initial configuration of \(f_n(x)\) was given by 
\begin{equation}
f_0(x) = \left\{
\begin{array}{cl}
0                                                         & x \in [a, 1)\\
1 - \frac{2}{(1-\epsilon)a} |x - \frac{(1-\epsilon)a}{2}| & x \in [0.0, (1-\epsilon)a)\\
1                                                         & x \in [(1-\epsilon)a, a), 1
\end{array}
\right.
\end{equation}
With the evolution of our functional dynamics, the function
\(f(x) \) for \(x \in I^2_0, I^2_1\) generates the Nagumo-Sato map.
The remaining part ($I_{rest}$) of the interval (i.e., that which is mapped according to a  distorted tent map) folds by
itself (see I) 
and if it is mapped to a value in $I^2_0$ or $I^2_1$,
it subsequently evolves under the generated Nagumo-Sato map. 
Figure \ref{fig:sato3}(b) shows snapshots of the function $f_n(x)$ for \(n = 100, 101\).
The function converges to a periodic function as a whole.
The period of the cycle is derived from the generated map.
The functional values of two different points $x'$ and $x''$ having the common periodic cycle changes
synchronously ($f_n(x')=f_n(x'')$), because 
the difference in $f_n(x)$ values decreases during the transient process before 
$f_n(x')$ and $f_n(x'')$ are attracted to the periodic motion, 
and also the Nagumo-Sato map has a contraction
property (with slope less than 1), in each branch.
As $n \rightarrow \infty$, the points in the interval $I$ are contained in either $I^1, I^2$ or $I^3$, and $I_{rest}$ vanishes.
\\

In general, an $f(x)$ with (at least) two type-I fixed points has 
a potential to possess a Nagumo-Sato map as a generated map.
To consider a general situation with multiple type-I points and/or
with several type-II intervals, we define the `multi-branch Nagumo-Sato map' 
by (6). In this case, $g(x)$ has the same slope $(1 - \epsilon) < 1$ for 
all $x$.
This type of map can be generated generally from random initial conditions.
In fact, in the inset in Fig.\ref{retR}(b), $g(x)$ consists of several branches
with the same slope $1 - \epsilon$.

With the multi-branch Nagumo-Sato map,
a function $f_n(x)$ periodic in $n$ with an arbitrary period can exist
for all $\epsilon$.
If $g(x)$ with a period-$m$ attractor is given, 
we denote a value of a type-III $f_n(x')$ as $a_{n+1} = g(a_n), n\bmod m$. 
Then a new attractor with period-$m+1$
is obtained by choosing an initial function to have two new branches properly.
We can arrange branches $[i]$ and $[j]$ to satisfy the
conditions that $g_n[i](a_{m-1}) = a_m$ and $g_n[j](a_m) = a_0$
for an arbitrary periodic orbit.\footnotemark
\footnotetext{There is some
restriction on  $a_m$ so that $x_i$ and $x_j$ cannot be the same.}

By choosing initial functions suitably, we can have
rather complex dynamics based on the multi-branch Nagumo-Sato map.
In Appendix B, the coexistence of multiple attractors is demonstrated,
while it is also shown that
$g(x)$ can have countably infinite attractors by suitably
choosing the initial conditions to generate the
multi-branch Nagumo-Sato map.

In the argument above,
the function $g(x)$ is defined at a countable number of points of $x$ 
(i.e., the attractor of the generated map $x_{n+1} = g(x_n)$ has a measure zero basin).
However, if all type-II fixed points are in a continuous type-II fixed 
interval, each attractor has a finite measure basin.
When $f_0(x)$ is a random function, such continuous intervals are formed.
In fact, a multi-branch Nagumo-Sato map is often generated
from random or other initial functions.  In general, the width 
of each branch is not identical, and a complicated combination
of branches is generated.  As shown in I, intervals of type-II fixed points
form a fractal structure.  Hence, branches in the generated map have 
an infinite number of
segments with a fractal configuration, in general.  Thus $f_n(x)$ 
can evolve with a complicated cycle that may be of 
infinite period.

\subsection{Case with Continuous Type-I Fixed Intervals}

In the cases considered to this point, the generated map in the functional dynamics (1)
cannot exhibit chaotic instability, in the sense that the
slope of the map is less than 1 for almost all points.
Except for a (countable) set of discontinuous points,
all generated maps have a slope \(1 - \epsilon\).
Here we study how a generated map can have
a larger class of 1-dimensional maps that allow for
chaotic instability.

To study this class of functional dynamics,
we extend our consideration to the case with
a continuous set of type-I fixed points, i.e., with
an interval of  type-I fixed points ($f(x') = x'$ for all  $x'\in I^1_i$).
The existence of such an interval is exceptional in
this functional map system, in the sense that 
it is almost impossible to produce such an interval  by the evolution (1)
unless the initial function does not include such an interval.
Indeed, a monotonically increasing function converges to a step function, and 
a single-humped function tends to converge to a function consisting of
isolated type-I fixed points and continuous intervals of 
type-II fixed points\cite{I}.
%Universality of this type of initial functions is discussed in Sec.5 and appendix A.

Although an initial function evolving into a function possessing type-I fixed intervals is 
rather rare in functional space, such an initial function may
have some meaning in our model.  The region in which $f(x)=x$ is nothing but a region
where the input is accepted `as it is'.  With regard to language, it is not absurd to assume that
some external input is transferred
without being modified or articulated.  We can consider that the
initial intervals satisfying $f_0(x) = x$ represent regions corresponding to such external input.
Other regions of $f(x)$ that are mapped to this part process
this external input according to the functional dynamics.
In addition to this meaning, the situation in which there exists a continuous set of type-I fixed
points is convenient for studying the hierarchy of meta maps, to be 
discussed in the next section.
Accordingly, we assume the existence of type-I fixed intervals.

Then, we define sets of type-II fixed points in
the same way as in the last subsection.
The type-I fixed intervals are labeled as $I^1_0, I^1_2, \ldots, I^1_{n-1}$.
Now, $I^2_i$ is defined as an interval where $f(x') \in I^1_i$ 
for all $x' \in I^2_i$ (See Fig.\ref{fig:ma}).
Although in the last subsection, we considered the case in which $f(x)$ is a 
constant function over an interval $I^2_i$, in the present case, 
each $f(I^2_i)$ can have various values 
in the range of $I^1_i$.
Let us write $f(I^2_i)$ as $f[i](x)$.
The generated map from the interval $I_{map}$ with fixed function has 
the
form

\begin{equation}
g(x) = 
\begin{array}{cccc}
\cup_i g[i](x) = & \cup_i (1 - \epsilon)(x - f[i](x)) + f[i](x)  & x \in I^2_i, & f[i](I^2_i) \subset I^1_i\\
\end{array}
\end{equation}
This function $g[i](x)$ is bounded both from above and below, 
because $f[i](x)$ has a possible minimum value $\min I^1_i = a$ and 
possible maximum value $\max I^1_i = b$,
$(1 - \epsilon)(x - a) + a \le g[i](x) \le (1 - \epsilon)(x - b) + b$ (see Fig.\ref{fig:ma}).
It is natural to call this function $g[i](x)$ within this bounded area the
`$i$-branch', in analogy to the last subsection.
For each type-I fixed point $x_i$, 
the generated map is given by 
$(1 - \epsilon)(x - x_i) + x_i$, although $x_i$ here can change continuously.

Consider the union of $n$ type-I fixed intervals 
$I^1 = I^1_0\sqcup I^1_2\sqcup \cdots\sqcup I^1_{n-1}$.
If type-II fixed intervals are within an interval $(\min I^1, \max I^1)$
and the condition $g(I^1 \sqcup I^2) \subset I^1 \sqcup I^2$ is satisfied,
the interval $I^1\sqcup I^2$
is a CGM (unit-I).
Here, in a type-II fixed point interval 
corresponding to a type-I fixed point interval, the generated map
\(g(x) = (1-\epsilon)x + \epsilon f(x)\) no longer has a constant slope. 
Rather the slope $g'(x) = (1 - \epsilon) + \epsilon f'(x)$ varies with $x$.

Following the argument in the last section,
we start from the case with two type-I intervals.
Now, we divide the interval \(I\) into a type-I fixed interval $I^1$ and
a type-II fixed interval $I^2$ ($I_{map} = I^1\sqcup I^2$).
Then, the interval \(I^1\) is divided into two parts, \(I^1_0\) and \(I^1_1\).
Without loss of generality, we can take $\min I^1_0 = 0$ and $\max I^1_1 = 1$.
The fixed function consisting of type-II fixed points is determined as 
\(f[0](x) \in I^1_0\) for \(x \in I^2_0\)
or \(f[1](x) \in I^1_1\) for \(x \in I^2_1\).
Then the  generated map \(g(x)\) is given by
\begin{equation}
g(x) = \left\{
\begin{array}{cl}
x                               & x \in I^1\\
g[0](x) = (1-\epsilon)x + \epsilon f[0](x) & x \in I^2_0\\
g[1](x) = (1-\epsilon)x + \epsilon f[1](x) & x \in I^2_1
\end{array}
\right.
\end{equation}
The area where the generated map can exist is denoted by the dotted area 
in Fig.\ref{fig:ma}.
Any 1-dimensional map included within the dotted area can be 
embedded into our functional map
by choosing the configuration of the type-II fixed function within the
shadowed area.
Since any function can be embedded in the dotted area, it is possible 
to have a case with $|g'(x)| > 1$.

In Fig.\ref{fig:ma}, there are regions where $g(x) \in I^1$.
If the generated map exists in such a region,
a point evolving as a type-III point may be absorbed into this
region and become a type-II point. Indeed,
when the point is mapped into this region, $f_{n+1}(x') \in I^1$ is satisfied,
and the point $x'$ becomes a type-II fixed point.
On the other hand, if $g(x) \in I^2$, a type-III point remains a
type-III during the entire evolution, and it never becomes a type-II fixed point.\\

Is there some restriction on the possible form of a generated map
allowed by the present functional dynamics?  As
discussed in Appendix C, there is some restriction according to
the present embedding of the generated map.  However, 
as is also discussed in that Appendix,
an arbitrary 1-dimensional map can be embedded as a generated map
by considering a two-step iteration, i.e., as a map to generate $f_{n+2}(x)$
from $f_n(x)$.\\

\section{Meta-Map in Functional Dynamics}

In Sec.3, we have shown how a 1-dimensional generated map is 
formed by a suitable configuration of type-I and type-II fixed points.
In the example, the 1-dimensional map is explicitly constructed 
with the condition $g(I^1 \sqcup I^2) \subset I^1 \sqcup I^2$. 
We call the interval $U^1\sqcup U^2$ a `unit-I' $U^1$.  In this case the generated map
$g_n(x)$ is fixed in time.
However, the CGM condition ($g(J) \subset J $) does not necessarily impose
the condition for the `unit-I'.  Then, $g_n(x)$ is not necessarily a fixed function.
In this section we consider such case in which a generated map changes dynamically in time.

In the situation discussed in Sec.3, 
in order for there to exist a generated map to determine the dynamics of the type-III points, it is
essential that the map stays within a bounded area.
The type-I fixed intervals are areas where type-II fixed points can exist,
from which the type-II fixed function never leaves.
The configuration of type-I and II fixed points determines bounded areas in which
the generated map remains as a branch.
The type-II fixed function determines a generated map within the bounded areas (See Fig.\ref{fig:ma}).

In the last section, we considered the situation in which the dynamics of the type-III point 
determined by $g(x)$ evolves within the interval $U^1$,
according to the type-II and corresponding type-I fixed points.
This process 
can be extended hierarchically.
In this section, we consider a unit-I instead of a type-I fixed interval 
and a type-III point instead of a type-II point, to see 
the dynamics of $f_n(x)$ for $x$ determined by the type-III point.

The unit-I ($\in U^1$) determines an interval where type-II and III points can exist.
The unit $U^1$ is $I^1 \sqcup I^2$.
Thus, $\{f_n(x)| f_n(x) \in U^1, x\in I^2\sqcup I^3\}$ consists entirely of type-II fixed points
and type-III points.  Here, the dynamics of the type-III points are determined by a CGM, and
their motion is confined within this region.
Thus, we replace the type-I fixed interval with unit-I 
by the transformation (5).  In case considered in the last section, the configuration of
type-I fixed intervals determines where the branches exist.
Here, the arrangement of unit-I determines where the branches of the
generated map exist.

First, we elucidate the branch structure determined by
a unit-I (See.Fig.\ref{fig:map.h2}).
These branches are derived from type-I and II points.
As noted, at the branch derived from a type-I point,
a 1-dimensional map is given according to the configuration of the type-II 
fixed function.  In the same way,
at a branch derived from type-II points,
a bounded map $g_n(x)$ exists according to the configuration of the type-III function.
Since $\{f_n(x)| x\in I^3\}$, consisting of the type-III points, depends on $n$,
the generated map, $g_n(x) = (1-\epsilon)x + \epsilon f_n(x)$,
also depends on $n$:

\begin{equation}
\begin{array}{l}
\left\{
\begin{array}{cll}
f(x)   &                                      & x \in I^2, f(x) \in I^1\\
g(x)   &= (1 - \epsilon)x + \epsilon f(x)       & x \in I^2\\
\end{array}
\right.\\
\left\{
\begin{array}{cll}
f_n(x) &= g(f_{n-1}(x))                       & x \in I^3, f_n(x) \in I^2\\
g_n(x) &= (1 - \epsilon)x + \epsilon f_n(x)   & x \in I^3  
\end{array}
\right.
\end{array}
\end{equation}
Here, $g_n(x)$ is generated from the type-III function and change with time step $n$.
The point $f_n(x') \in I^3$ evolves according to $g_n(x)$ ($f_{n+1}(x') = g_n(f_n(x'))$).
Since $g_n(x)$ is not a fixed function, 
it represents a change of rules. Accordingly,
we call this type of map a `meta-map'.
By using these branches, we can construct a new CGM consisting of $g(x)$ and $g_n(x)$.
The type of point $x$, which evolves under the CGM, can change in time.
The interval $I$ can be written $I^1\sqcup I^2 \sqcup I^3_n \sqcup I^4_n \sqcup I_{rest}$.
Now, $I^3$ and $I^4$ have the suffix $n$.

When the dynamics of type-III points are periodic, determined by a (multi-branch)
Nagumo-Sato map, the dynamics of a type-IV point determined by the type-III 
points is also periodic.  Indeed, this hierarchical structure is
often formed starting from a random initial function, since a generated map of the Nagumo-Sato type
is commonly formed, as mentioned in Sec.3.
In Fig.\ref{fig:type-ch2}(a), an example of a meta-map (return map) 
is plotted. These data were obtained with a numerical simulation starting from
a random initial function (see Sec.3.1) $f_0(x)$.
For the points indicated by arrows, the return map has two values.  Hence the dynamics of the points
are not determined by a fixed generated map, but by a time-dependent generated map.
In this case, the `type' of a point $x'$ is no longer fixed, but can change between
type-IV and type-III, depending on the intervals in which $f_n(x')$ is situated as $n$ changes.
The evolution of the `type' of a particular $x'$ is plotted in Fig.\ref{fig:type-ch2}(b).

A simple example of the `type' change is displayed in Fig.\ref{fig:type-ch}.
Here, fixed points in the right-hand part generate $g(x)$,
which determines the dynamics of type-III point with period 2.
The type-III points generate a time-dependent map that switches between
$g_{odd}(x)$ and $g_{even}(x)$.
The fixed points on the left-hand side generate $g[0](x)$, which determines the
dynamics of the type-III points.  Here, 
the fixed map consisting of $g[0](x)$ and $g_{even}(x)$ generates a period-2 orbit.
If the evolution of the $f_n(x')$ is determined by $g_{even}(x)$ at even $n$  
or $g[0](x)$ at odd $n$, 
$f_n(x')$ changes cyclically with period 2 as 
\begin{equation}
\begin{array}{l}
\left\{
\begin{array}{ccl}
f_{even}(x') &=& g[0](f_{odd}(x'))\\
f_{odd}(x')  &=& g_{even}(f_{even}(x'))
\end{array}
\right.
\end{array}
\end{equation}
and its `type' also changes cyclically
with period 2 as III, IV, III, IV, $\cdots$.
On the other hand, if the evolution of $f_n(x')$ is determined by $g_{even}(x)$ at odd $n$
or $g[0](x)$ at even $n$, $f_n(x')$ is a type-II fixed point.\\

When a type-III point possesses a chaotic orbit, as given in 
Sec.3.3, the nature of the functional dynamics determined by this type-III point 
is more interesting.  Let us study the case with a chaotic generated map 
by constructing an example.
By choosing a suitable initial function,
one can embed a 1-dimensional map to construct a meta-map explicitly.
For example, we adopt the following 1-dimensional map to be embedded:
\begin{equation}
\begin{array}{cc}
x_{n+1} = 2(1 - \epsilon)x_n + \epsilon & (\bmod 1)
\end{array}
\end{equation} 
This map has chaotic orbits for $\epsilon < 1/2$.
Indeed, the initial function $f_0(x)$ 
\begin{equation}
\begin{array}{l}
f_0(x) = \left\{
\begin{array}{cl}
x                       & x \in [0, (1- \epsilon)/2], [(1+ \epsilon)/2, 1.0]\\
\frac{1 - \epsilon}{\epsilon }x + (3 \epsilon -1)/(2 \epsilon)  & x \in ((1- \epsilon)/2, 1/2]\\
\frac{1 - \epsilon}{\epsilon }x - (1 - \epsilon)/(2 \epsilon)      & x \in (1/2, (1+ \epsilon)/2)
\end{array}
\right.
\end{array}
\end{equation}
generates this map.
In Fig.\ref{fig:meta-map0}(a), $f_0(x)$ is plotted for $\epsilon = 1/3$.
The dotted line represents a function $g_0(x)$ that has exactly the form $x_{n+1} = 2(1 - \epsilon)x_n + \epsilon$.

To construct a meta-map, we replace the type-I fixed interval in Fig.\ref{fig:meta-map0}(b) 
with this $f_0(x)$ (unit-I) by the transformation (5).
With this nested structure, there are intervals where a type-III function 
exists, and the function generates an $n$-dependent $g_n(x)$ (CGM).
This CGM acts as the map for a point that satisfies
$f_n(x') \in I^1\sqcup I^2\sqcup I^3$.
For this configuration, the dynamics of a part of a CGM are determined 
by type-III points. In this case, a point which satisfies $f_n(x') \subset I^3$ 
behaves as a type-IV point under the iteration.

In Figs.\ref{fig:meta-map}(a)-(e), the evolutions of $f_n(x)$ and the 
meta-map are plotted, where the map $g_n(x)$ is plotted as a dotted line.
Type-III points evolve according to  
$f_{n+1}(x') = 2(1 - \epsilon)f_n(x') + \epsilon$.
In the present example, the slope $g'(x) = 2(1 - \epsilon)$ is constant, 
and the slope of the type-III function $(f_n(x))$ is easily calculated as:
$f'_n(x) = \frac{(1 - \epsilon)}{\epsilon} 2^n(1- \epsilon)^n$.
The generated map is determined as $g'_n(x) = (1 - \epsilon) + \epsilon f'_n(x)$, 
and it has the form
$g'_n(x) = (1 - \epsilon)\{1 + 2^n(1 - \epsilon)^n)\}$.
Hence, this meta-map has a part where its gradient increases 
exponentially with $n$.

This implies that our functional dynamics can have stronger orbital
instability than deterministic chaos:  
%The difference of $g'_n(x)$, 
%$\Delta g'_n(x) = g'_{n+1}(x) - g'_n(x) = 
%(1- \epsilon)(1 - 2\epsilon)2^n(1- \epsilon)^n$.
%If $\epsilon < 1/2$, $\Delta g_n(x)' = c\alpha ^n$ 
%with $c$ as a constant and $\alpha > 1$.  
A tiny deviation $\delta$ from a point mapped to this type-III points grows as 
$\prod_{n = i}^{N}|g'_n(x)|$.
Since $\prod_{k = 1}^{n} \alpha^k = \alpha^{\frac{n(n+1)}{2}}$, the leading order 
of the exponent of the orbital instability is $n^2$.
Hence, the orbital instability is such that a tiny deviation grows as 
$exp(const \times n^2)$ rather than
$exp(const \times n)$, as is the case in conventional chaos.  
Due to this strong instability based on 
chaotic dynamics in the generated map, we call this dynamics 
`meta-chaos'.
In Fig.\ref{fig:meta-map}(f), an example of the orbits for meta-chaos
is displayed. This evolution is determined by $g_n(x)$.
The `type' of the point changes between III and IV according to the map $g_n(x)$.
%directed by the point.

For a numerical simulation with this meta-chaos, the required mesh size
increases as $2^n$.  Hence, a simulation quickly
becomes invalid as $n$ increases.
%In the example of Fig.\ref{fig:meta-map}(f), the function 
%never goes out the defined interval and 
%the transition sequence of types between III and IV can be computed
%for longer steps correctly, while the orbit of the function itself departs
%from the correct one within shorter steps.  

In the example mentioned above, we have constructed a meta-map by choosing
special initial conditions.  However, we  
note again that a
meta-map configuration itself is not special and can be reached, for
example, from a random initial function.
Still, it is very rare to obtain a continuous type-I fixed interval from random
initial conditions.  Hence, in most simulations from arbitrarily chosen initial
functions, we mostly observe generated maps of the Nagumo-Sato type, where 
magnitude of the slope, $|g_n'(x)|$, is always less than 1.\\

The nesting process of the meta-map can be continued hierarchically, since
the configuration of type-I, II and III points, discussed above, can
be a CS, which generates a CGM as a whole.
This arrangement is called a `unit-II'.
One can replace  a unit-I in the above construction by such a
unit-II. In such a situation, type-IV points generate a map, and with an appropriate
configuration, the generated map can be a CGM.
Now we can call this CGM a `unit-III'.
This hierarchy to form a `unit-$N$' can be continued for
$N \rightarrow \infty$ (see Fig.\ref{fig:sc2}).
To continue this nesting process, we define the `unit-$N$' and the $N$-th level meta-map as follows.
   
A unit-$N$ is an interval $U^N$ which consists of type-I, II,$\ldots$, $N+1$ points and
satisfies the condition $g(U^N) \subset U^N$.
A point $f_n(x') \in U^N$ (with $x' \not \in U^N$) evolves by the unit-$N$ and has a `type'
from I to $N+2$.
We denote a function defined on an interval of type-$N$ points as $f^N_n(x)$ 
and the generated map from $f^N_n(x)$ as $g^N_n(x)$ (a fixed function is written $f(x)$ instead of $f^{II}(x)$).
The functional equation rewritten in a recursive form with respect to the `type' has the form
\begin{equation}
\left\{
\begin{array}{cll}
f^N_n(x)   &= g^{N-1}_n(f^N_{n-1}(x))               & x \in I^N, f(x) \in I^{N-1}\\
g^N_n(x)   &= (1 - \epsilon)x + \epsilon f^N(x)       & x \in I^N
\end{array}
\right.
\end{equation}
The $N$-th level meta-map is defined as a CGM consisting of 
$g(I^1\sqcup I^2), g^{III}(I^3_n), \ldots, g^{N+2}(I^{N+2}_n)$, where
a 1-dimensional map generated from type-I and II points is called the `0-th level' meta-map.
All meta-maps depend on the fixed function $f(x)$ and are constructed recursively as
$f(I^1 \sqcup I^2), f^{III}(I^3_n), \ldots, f^N(I^N_n)$.
The whole interval $I$ can be written 
$I^1\sqcup I^2\sqcup I^3_n \sqcup \cdots \sqcup I^N_n \sqcup \cdots$.
Here note that a `type' greater than 2 can change in time,
although each point has a finite maximal value of its `type', depending on the initial configuration.

The $N$-th level meta-map is determined by the configuration of type-I, II, $\ldots, N+2$ points.
It is important that each unit-$N$ and each branch are bounded.
We can arbitrarily arrange any unit-$N$ and type-III, IV, 
$\ldots N+2$ points according to the branches.
The configuration producing a meta-map characterizes a `syntax' for each $x$.
Each $x$ has a time evolution as a type.
The `type' of a point that is of type-III or higher changes in time.
For a meta-map higher than second level, 
there is a sequence, for example, III, III, IV, V, III, $\ldots$.
There is a transition relation among type-$N$ ($N > 2$) points.
Each point evolves under a hierarchy of meta-maps.
In the above representation, the dynamics of the $N$-th level meta-map is independent of that of the 
type-$N+3$ points.

In a high level meta-map, the orbital instability is stronger than the exponential instability of conventional chaos.
If $g'(x) \sim \alpha$ $(|\alpha| > 1)$ and $f^{III}_i(x)$ 
with a gradient $\beta$ is not a constant function (i.e., $\beta \neq 0$), 
then $f^{III\prime}_n(x) \sim \beta\prod_{k = i}^{n} \alpha \sim \alpha^n$, and $g^{III\prime}_n(x) \sim \alpha^n$.
Now, the leading order of the slope of a 1st-level meta-map is $\alpha^n$, as is mentioned above.
A type-IV function $f^{IV}_n(x)$ evolves under this 1st level meta-map.
If $f^{IV}_i(x)$ is not a constant function and has a gradient $\gamma \neq 0$, $f^{IV\prime}_n(x)$ is calculated as 
$\gamma \prod_{k = i}^{n} \alpha^k \sim \alpha^{n^2}$, and $g^{IV\prime}_n(x) \sim \alpha^{n^2}$.
Hence, $f^{V\prime}_n(x) \sim \alpha^{n^3}$ and $g^{V\prime}_n(x) \sim \alpha^{n^3}$.
Repeating this argument, 
the leading order of the slope of the $N$-th level meta-map is given by $\alpha^{n^N}$.
Thus a tiny deviation from a point, which evolves under the meta-map, is amplified by $|\alpha|^{n^N}$ at each $n$ step.
Because of this, an $N$-th level meta map has an orbital instability that behaves as $exp(const \times n^{N+1})$.
The level of the orbital instability increases with the level as $exp(const \times n^{N+1})$.
In other words, an exponent $\lambda$ corresponding to the Lyapunov exponent of conventional chaos increases as
$\lambda \sim n^N$ as $n$ increases for the $N$-th level meta-map.

\section{Summary and Discussion}

In the present paper, we have studied functional dynamics, focusing
on the generation of rules (mappings) for the dynamics representing change
of a function, and on the hierarchy of meta-rules. 

As a first step, we introduced a new concept, the `generated map' $g(x)$,
which is derived from $f_n(x)$ and determines the dynamics of $f_n(x)$.  The dynamics of some other parts of $x$
are determined by this generated map, while a closed generated map is defined as
one that maps a region into itself.
Functional values on some intervals were shown to change according to the
generated map. This leads to a
1-dimensional map or a `meta-map' that changes the map itself.

In Sec.3, we explicitly showed that some classes of 1-dimensional maps
are embedded into this functional dynamics.
In Sec.3.1 and 3.2, a piecewise linear map with two intervals of the slope
$1-\epsilon$ were shown to be generated from two type-I fixed points and 
two intervals of corresponding type-II fixed points.
Next, this construction was generalized to 
cover the case with several isolated type-I fixed points and the corresponding
type-II intervals.  There,
a piecewise linear map with several intervals with slope $1-\epsilon$
were found to be generated. This map, called a `multi-branch Nagumo-Sato Map' exhibits 
periodic cycles.  Hence, 
the dynamics of the functional values determined by this generated map display a periodic
cycle, which explains why periodic cycles are often
generated in our functional dynamics.

In Sec.3.3, generated maps with continuous type-I fixed intervals and type-II 
points were discussed.  In this case, a 1-dimensional map with
an arbitrary slope can be embedded. Now, the functional dynamics determined by this
generated map can also exhibit chaotic dynamics.

As shown in Sec.4, this construction of generated maps can continue  
hierarchically.  The dynamics determined by a generated map 
forms a higher-level generated map that determines the dynamics of other regions.
Since this map is changed by the first generated map, 
it is regarded as a `meta-map',  a map determined by another map.
This procedure can be continued ad infinitum, leading to meta-meta-... maps.
When a generated map exhibits chaotic dynamics, as discussed in Sec.3.3, the
dynamics by meta-map can exhibit `meta-chaos', in the sense that the evolution rule
itself changes chaotically in time.  It was shown that this meta-chaos has
a stronger orbital instability than in chaos, in the sense that a small
deviation is amplified as $exp(const\times n^{M+1}))$ for the $M$-th level meta-map,
rather than $exp (const \times n)$.\\

Now, we discuss the relevance of our results for the
target problems listed in Sec.1. 
The basic structure of the functional dynamics is provided by
two types of fixed points, while a 1-dimensional map is generated 
by the configurations of fixed points.
The evolution of the type-III points is determined by the map generated by 
several type-II intervals.
A fixed point is invariant under iterations of the map
and can be regarded as the basis (words) for the description
of the world.  
The network of fixed points (word) generates the map $g(x)$ (rule).
A set of orbits of type-III points can be regarded as the abstraction and categorization for the words.
The words generate a rule, and the orbit determined by this rule determines a set of words.
Thus, there is a circulation between words and rules, 
in the sense that the network of words generates a rule and that an orbit determined by the 1-dimensional map can be 
regarded as a representation of the operation on the fixed point (word).

Similarly, the 1st-level meta-map determines an orbit consisting of type-III and IV points,
as an operation on a set of type-II and type-III points.
In this hierarchical configuration, each orbit is characterized by a sequence 
of types and a sequence of values $f_n(x)$.
A point, which evolves under the $N$th level meta-map, changes its ``types' variously, 
type-III, IV, $\ldots, N+3$.
The sequence of `types' determines the operation on elements 
which belong to various types in the hierarchy.
This is regarded as a representation of the modality of a connection between 
words/rules and rules/meta-rules.

A map and a meta-map determine an orbit, which evolves following a
lower level structure in the hierarchy.
In our system, a higher-level structure is formed based on 
the lower-level structure,
which we believe is an important characteristic in language.
For example, syntax with regard to words is generated
in our system, while this syntax depends on the words themselves.
The `type' of a word never changes in time,
and the operation (orbit) on the words is determined by the time evolution.

The articulation as a whole is represented as $I = I^1\sqcup I^2\sqcup I^3_n\sqcup \cdots \sqcup I^N_n\sqcup \cdots$.
The suffix $n$ is derived from the orbit of each `type'.
A possible region where an orbit can exist is determined by a branch of the generated map,
while the possible region where the branches can exist is determined by the orbit.
The abstract language/object space is organized by the real orbit, which is 
one of the possible orbits under the restriction that of the partition $A_n = \sqcup_i\{I^i_n\}$.

When a partition $A_n$ is given, an infinite variety of possible orbits can exist.
Depending on the evolution, a different $A_{n+1}$ can be formed.
%There is an order among sets $A$, the order is decided as $A_n \sqsubseteq A_{n+ 1}$.
%The real orbits which can exist in $A_0$ is infinite and there is a network with the order.
The syntax given by the partition $A_n$ crucially depends on the
orbit, which on the other hand is determined by the generated map
organized through $A_n$.  The syntax (i.e., the rule derived from the generated maps)
and the semantics (i.e, the orbit of functional dynamics) determine each other,
and cannot be derived independently.
This situation can be regarded as representing the `speech act', 
in which syntactic structure (a generated map) generates an utterance (orbit), 
and the utterance changes the syntactic structure.

In formal language theory, a grammar is
implemented independent of semantics, as an innate structure.
In our functional dynamics, such rules governing the use of  words 
are thought of as being formed only through iterations.  No other assumption except for
the choice of  non-trivial initial functions is required.  Since iterations are 
essential to language, we expect that the formation of  rules in our system, 
depending on words themselves, is relevant to the origin of syntax and semantics
in language.  

In our system, a hierarchical structure is also formed through iterations.
This hierarchy is also a characteristic of language,
and it is important to note that a simple class of functional dynamics
with recursive structure can provide such hierarchy in general.
The hierarchical structure in this system has a strong dependence on the
lower-level structure, since
the higher-level structure is determined according to which branch of the generated map
is taken by the orbit.

The form of the generated map $g(x)$ depends on
the configuration of type-I fixed points.  If they are
discrete, the slope of $g(x)$ is smaller than 1.
When there exists a continuous interval of fixed points,
$g_n(x)$ can have a slope larger than 1,
and the meta-map can have a more complex orbit than in chaos.
A continuous type-I fixed interval is generated by  
an identity function over some interval, which 
corresponds to a filter with which
an agent acts in response to the world without interpretation.
In other words, chaotic functional dynamics and meta-chaos are
generated by adding a continuous input from the external world
to the `closed' world of functional dynamics only with self-reference.

Possible extensions of the present study will be discussed in the future.
In a 2-dimensional version of the functional dynamics,
an arbitrary 2-dimensional map can be embedded in the same way as in Sec.3.3.
Because of this, we can embed a Turing machine into this system \cite{Moore} 
(see also Appendix B), where
the search for a relationship between the generalized shift \cite{Moore} and meta-dynamics (meta-chaos) 
will be important.

Non-trivial sets of functions over functions are studied in domain theory \cite{Plotkin} 
\cite{Varela} \cite{Rosen}.
The most important difference between systems studied in domain theory and our model
lies in the dynamical aspects of functions treated only in our approach.
However, our meta-map is restricted within some intervals and is not extended over the whole domain.
Indeed, in our system the size of the $N$-th level meta-map decreases with order $\epsilon^N$.
However, such a contraction can be removed in a more general functional dynamics.
This will be important to obtain functional dynamics allowing for a hierarchy of the meta-map over the whole domain.

Another extension required for language will be
the inclusion of dialogue \cite{III}.
To this point, we have only considered one agent whose function
changes recursively.  To study the social structure of language,
functional dynamics with several agents is necessary.

\clearpage
\appendix

\section{Some Properties of $F(x, y)$}

In this appendix, we investigate a general class of
functional maps with the form
\begin{equation}
f_{n+1}(x) = F(f_n(x), f_n \circ f_n(x)).
\end{equation}
We study a fixed point condition and  properties of the generated map.

This type of functional equation has fixed points (fixed functions).
First, we define $Z(x)$ from $F(x, y)$.
Here, $Z(x)$ is the solution of $x = F(x, Z(x))$.
The fixed point condition is defined from $Z(x)$.
If the condition $f \circ f (x') = Z(f(x'))$ is satisfied, $f(x')$ is a fixed point.
If $F(x, y) = (1 - \epsilon )x + \epsilon y$, then $Z(x) = x$, and the
fixed point condition is nothing but $f\circ f(x') = f(x')$.
The fixed point condition in the present general case is determined as follows.
(We give the correspondent equation for the case
with $F(x, y) = (1 - \epsilon)x + \epsilon y$ in the square bracket $[...]$, for reference.)

\begin{description}
\item[(i)]
If $Z(x)$ is a single valued function, $f(x) = Z(x)$ is a fixed function over the entire interval
($f\circ f(x') = Z(f(x')$).

[$f(x) = x$ is a fixed function]

\item[(ii)]
The point where $Z(x)$ intersects the identity function ($x' = Z(x') = f(x')$) is a fixed point
($f \circ f(x') = f(x') = Z(x') = Z(f(x'))$).

[Type-I fixed point condition]

\item[(iii)]
If a point $(x', f(x'))$ is a fixed point ($f(x') = Z(x')$),
a point $(x'', f(x''))$ which satisfies $f(x'') = Z(x'') = x'$ is also a fixed point
($f\circ f(x'') = f(x') = Z(x') = Z(f(x''))$).

[There is no such fixed point corresponding to this case]

\item[(iv)]
If a point $(x', f(x'))$ is a fixed point, a point $x''$ with
$f(x'') = f(x'))$ is also a fixed point.

[type-II fixed point]

\end{description}

The most noteworthy difference from
the case with $F(x, y) = (1 - \epsilon)x + \epsilon y$ is seen in (iii).
For a point $f(x'') = Z(x'')$, the fixed point condition is that $Z\circ Z(x')$ is a fixed point.
There, $Z(x)$ decides a fixed point condition as an orbit of a 1-dimensional map.
In other words, the `attractor' of $Z(x)$ is a fixed point of the equation (15), and
a sequence $\{f(x'), Z \circ f(x') = f(f(x')),
Z^2 \circ f(x') = f(Z \circ f(x')), \ldots, Z^{\infty} \circ f(x') \in $ attractor $\}$
consists of fixed points.

The functional equation can be divided into
\begin{equation}
\left\{
\begin{array}{lcl}
f_{n+1}(x) & = & g_n \circ f_n(x)\\
g_n(x)     & = & F(x, f_n(x))
\end{array}
\right.
\end{equation}
as in the case $F(x, y) = (1- \epsilon)x + \epsilon y$.
The generated map viewpoint is also effective in this general case.

However, for a general $F(x, y)$, the transformation (5) cannot be adopted,
because $F(x, y)$ is not linear.
However, the use of a generated map to construct a meta-map remains valid in a general $F(x,y)$ case,
and a hierarchical configuration can exist for a particular configuration.\\

\section{Multi-Branch Nagumo-Sato Map}

In general, $f(x)$ with (at least) two type-I fixed points has
the potential of possessing a Nagumo-Sato map as a generated map.
To consider the general situation, we define the `multi-branch Nagumo-Sato map'
by (6),  restricted within a region  $I = [x_0, x_{n-1}]$, while
$x_0, x_1, \ldots x_{n-1}$ can be arranged arbitrarily.
This type of map can be generated from random initial conditions.

In this map, we can choose a function which generates a map producing cycle of any
length of period. To illustrate this property, we study the case with
some special configurations.

First, two type-I fixed points are assumed to be 0 and 1.
For the sake of symmetry, we choose $\epsilon = 1/2$.  Then,
the two branches are given by
\begin{equation}
\left\{
\begin{array}{clc}
g[0](x) &= \frac{1}{2}x                       & x \in I^2_0, f(I^2_0) = 0\\
g[1](x) &= \frac{1}{2}x + \frac{1}{2}          & x \in I^2_1, f(I^2_1) = 1
\end{array}
\right.
\end{equation}
We represent a rational number $a$ by the binary form $0.a_1 a_2 a_3 \cdots$, with each $a_i = 0, 1$
(here $a = \sum_{k = 1}^{\infty} a_k 2^{-k}$).
In this representation, $g_0(x)$ acts as a right-shift, which acts as
$0.a_1 a_2 a_3 \cdots \rightarrow 0.0 a_1 a_2 a_3 \cdots$, and
$g_1(x)$ acts as a right-shift and inserts 1 into the head of the sequence as
$0.a_1 a_2 a_3 \cdots \rightarrow 0.1 a_1 a_2 a_3 \cdots$.
Hence these two branches act as 0, 1-inserter for a binary sequence.

Here we denote $a$ $ (a_i = a_{i+m})$ as $\{a_1 a_2 \cdots a_m\}$,
while the set of $m$-length sequences $\{a_1 a_2 \cdots a_m\}$
is denoted by $S_m$.
The number of elements which belong to $S_m$ is $2^{m-1}$, and
the values of $a$ $(\in S_m)$ take $i/2^{m-1}$ ($i = 0, 1, \ldots, 2^{m-1}$).
If $\{a_1 a_2 \cdots a_m\} \in S_m$, $\{a_m, a_1 a_2 \cdots a_{m-1}\} \in S_m$.
Because of this, when $g(\{a_1 a_2 \cdots a_m\}) \equiv g_{a_m}(\{a_1 a_2 \cdots a_m\})$,
the map $g(x)$ is a bijection $S_m \rightarrow S_m$.
We define $M_m(x) = g(x)$ over $S_m$.

$M_m(x)\cup M_n(x)$ is a single-valued function for arbitrary $m, n$.
The condition that a point $a \in S_m \cap S_n (m < n)$ exists is that $m$ is a divisor of $n$.
In such a case, $a$ has the form $\{a_1 a_2 \cdots a_m\} \in S_m$, and
$\{a_1 a_2 \cdots a_m\}^{n/m} \in S_n$.
These two representations determine the same $g(a)$.
Then, $M_{\infty}(x)$ defined as $\cup_{k=1}^{\infty} M_k(x)$
has an infinite period.

The function $M_m(x)$ is defined at $2^{m-1}$ points.
With an appropriate arrangement,  it is possible
for $g(x)$ generated by the attractor of our functional dynamics
to be made equal to $M_m(x)$ for all $x$.
As an example, we define g(x) as $g_{o(i)}$
for an interval $[i/2^{m-1}, i+1/2^{m-1})$ $(i = 0, 1, \cdots 2^{m-1}-1)$.
Here, $o(i) = 0$ for even $i$ and $o(i) = 1$  for odd $i$.
In Fig.\ref{fig:multi2}, we can take a section$_i$
$[i/2^{m-1}, (i+1)/2^{m-1}) \times [g(i/2^{m-1}), g(i/2^{m-1})+1/2^{m-1})$
within the region where $g(x)$ is defined. The map $g(x)$ determines the bijection section$_i \rightarrow$ section$_j$.
In each section, $g(x)$ has a slope 1/2 ($< 1$), and all orbits converge to attractors
which are determined by $M_m(x)$.
Thus we can embed a multi-branch Nagumo-Sato map which has multiple attractors.

In the same way, we can construct an $n$-branch Nagumo-Sato map.
We assume $x_i = i/n$ $i = 0, 1, \ldots, n-1$.
If $\epsilon = (n-1)/n$, each branch has the form
\begin{equation}
g[i](x) = \frac{1}{n}x + \frac{i}{n}          (x \in I^2_i, f(I^2_i) = \frac{i}{n})
\end{equation}
Each branch $g[i](x)$ indicates a right-shift and insertion of
$i$ at the head of the $n$-digit sequence.
Using these branches, we can embed an $m$-periodic point for
an $n$-digit representation $\{a_0, a_1, \ldots, a_{m-1}\}$ $(a_i = 0, 1, \ldots, n-1)$.\\

\section{Embedding a General One-Dimensional Map as a Generated Map}

\begin{figure}
%\begin{center}
\noindent
\epsfig{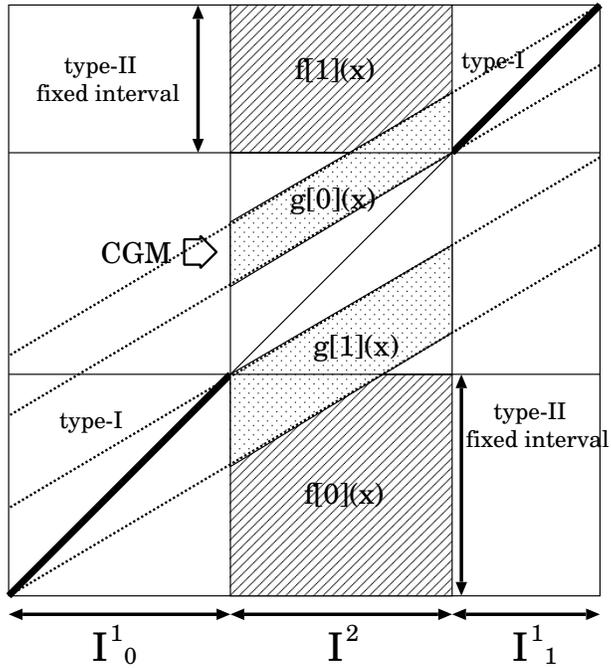}

\caption{The closed 1-dimensional map.
Using two type-I fixed intervals, 
we can obtain a larger class of a 1-dimensional map.
The type-II fixed function \(f(x)\) is represented by $f[0](x)$ (for $x \in I^1_0$) and
$f[1](x)$ (for $x \in I^1_1$). 
\(I^2\) is a domain of the type-II fixed function \(f(x)\).
The generated map \(g(x)\) is given by
$g[0](x) = (1 - \epsilon)x + \epsilon f_0(x)$  (for $x \in I^1_0)$ and
$g[1](x) = (1 - \epsilon)x + \epsilon f_1(x)$  (for $x \in I^1_1)$.
We call this type CGM a `unit-I'.
In this figure \(\epsilon = 0.5\).
}
\label{fig:ma}
%\end{center}
\end{figure}

\begin{figure}
%\begin{center}
\noindent
\epsfig{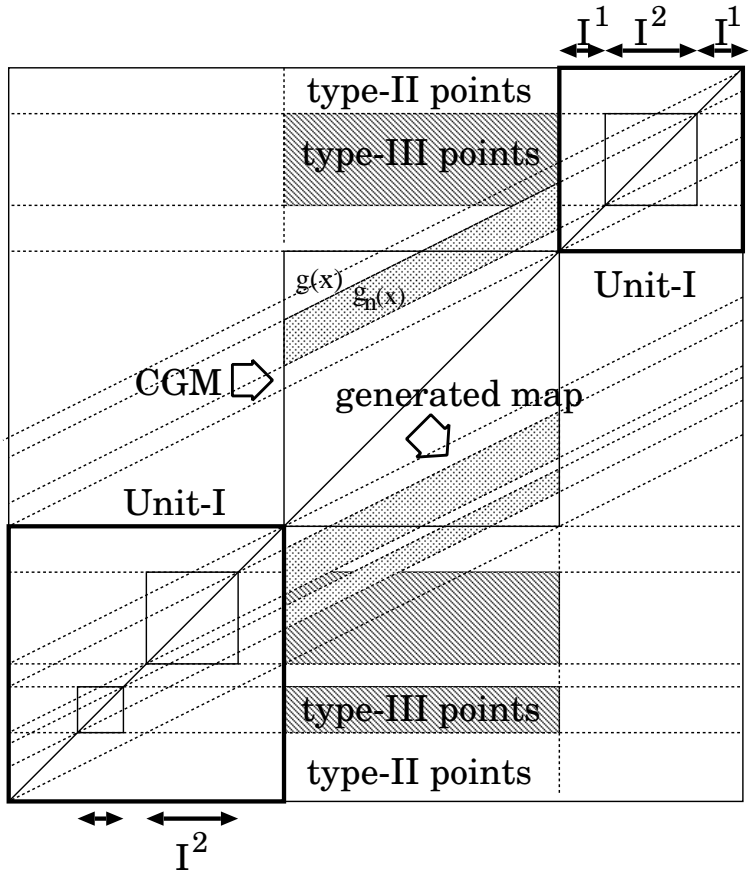}

\caption{The hierarchical arrangement of a `meta-map'.
The shaded area is a region where type-III points can exist and the dotted branch indicates 
a region where the generated map, derived from type-III points, can exist. 
A generated map in $I^3$ has an $n$ dependence ($g_n(x)$).
}
\label{fig:map.h2}
%\end{center}
\end{figure}

\begin{figure}
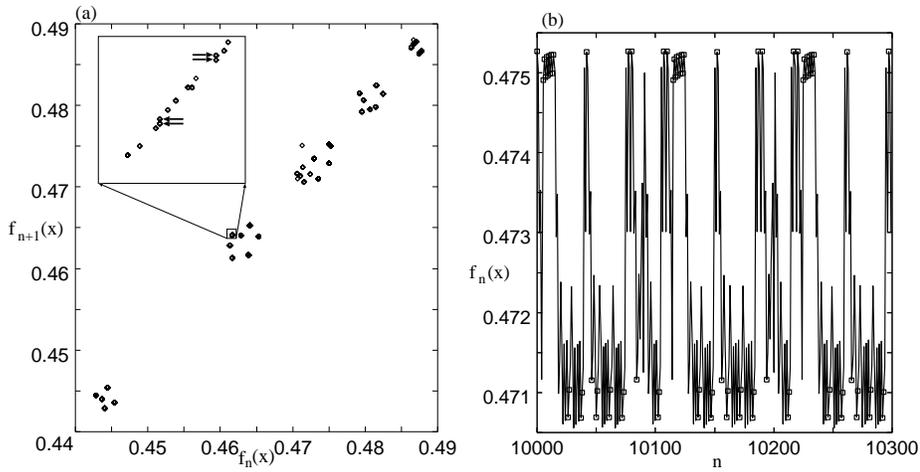

%\begin{center}
\noindent
\epsfig{file=met.5.eps,height=6cm,width=6cm}
\epsfig{file=V.eps,height=6cm,width=6cm}
\caption{(a) Another close up of the return map of Fig.1(a).
At the points $x'$ indicated by arrows, $g_n(x')$ has two values. 
(b) Time evolution of $f_n(x')$ for $10000 < n < 10300$, 
which is determined by the meta-map and has a period 111. 
The function $f_n(x')$ is a type-IV point at the steps $n$ plotted 
with squares $\Box$, while it is type-III otherwise.
}
\label{fig:type-ch2}
%\end{center}
\end{figure}

\begin{figure}
%\begin{center}
\noindent
\epsfig{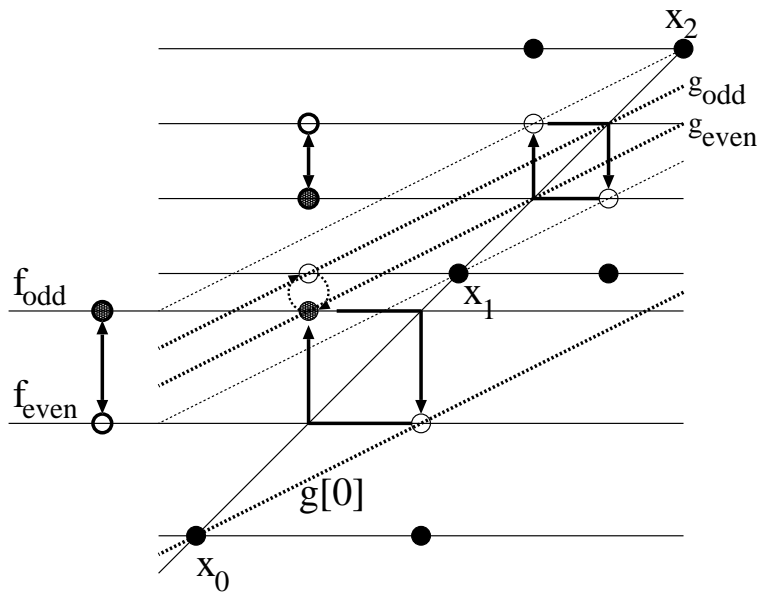}
\caption{One of the simplest configurations for `type' change.
Type-I fixed points $x_1, x_2$ and two type-II points generate a period-2 map (as in the upper-right square).
A period-2 motion determined by the generated map generates a period-2 branch $g_n(x)$.
Then, $g[0](x)$ and $g_{even}(x)$ produce a period-2 orbit (as in the lower left square).
If the evolution of a point $f(x')$ is determines by $g[0](x)$ ($g[even](x)$), $f(x')$ is a type-III (IV) point.
The sequence of the `type' of $f(x')$ is III, IV, III, IV, III, $\cdots$.
}
\label{fig:type-ch}
%\end{center}
\end{figure}

\begin{figure}
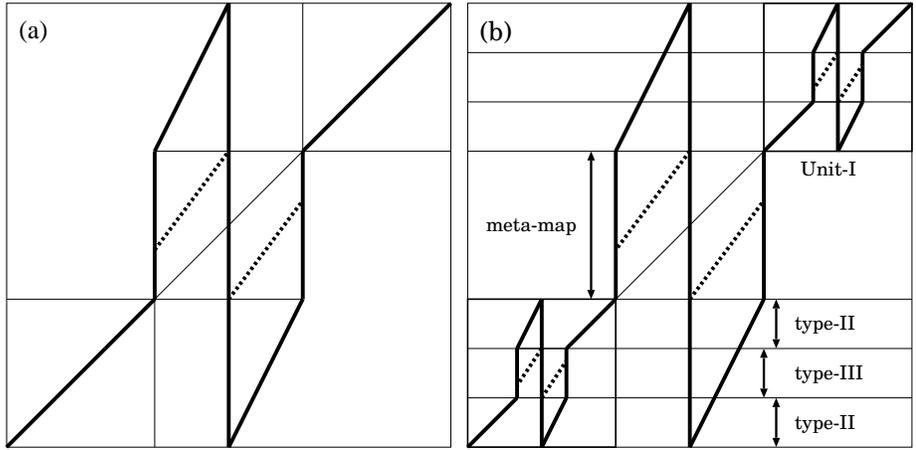

%\begin{center}
\noindent
\epsfig{file=meta1.eps,height=6cm,width=6cm}
\epsfig{file=meta2.eps,height=6cm,width=6cm}
\caption{(a) A configuration (indicated by the solid line) 
which generates a map $x_{n+1} = 2(1 - \epsilon)x_n + \epsilon$ $(\bmod 1)$ (indicated by the dotted line) 
in the center square ($\epsilon = 1/3$).
(b) An initial function $f_0(x)$ leading to a hierarchical configuration of the map (a). 
}
\label{fig:meta-map0}
%\end{center}
\end{figure}

Let us examine closely the configuration of the type-I fixed intervals
adopted to embed a 1-dimensional map.
The area in which a 1-dimensional map is embedded
has to be on the intersection between each branch of type-II intervals and
$I \times \bar{I}^1_i$ (See.Fig.\ref{fig:ma}).
This implies that we cannot embed a map which is continuous around the
identity function.
However, one can embed an arbitrary 1-dimensional map by considering a
two-step iteration, i.e., as a map to generate $f_{n+2}(x)$
from $f_n(x)$.
As shown in Fig.\ref{fig:logi}, let us
take two maps in the dotted areas of Fig.\ref{fig:ma}.
As shown in the figure, the generated maps \(g[0](x)\) and \(g[1](x)\) are put in
two regular square sections. Here,
\(f_n(x^{\prime})\) which is mapped to \(g[0](x)\) evolves as
\(f_{n+1}(x^{\prime}) = g[0](f_n(x^{\prime}))\) and
\(f_{n+2}(x^{\prime}) = g[1] \circ g[0](f_n(x^{\prime}))\).
If \(g[0](x)\) is the identity function,
the time evolution of \(f_n(x^{\prime})\) at \(n = 2i\) (\(i\) integer)
is \(f_{n+2}(x^{\prime}) = g[1](f_n(x^{\prime}))\), \(f_{n+4}(x^{\prime}) = g[1]
(f_{n+2}(x^{\prime}))\)
\(f_{n + 2i} = a^i(f_n(x^{\prime}))\).
Hence an arbitrary 1-dimensional map can be embedded as a rule for the
two-step iteration of the functional dynamics.\\

{\bf Acknowledgments}

The authors would like to thank Drs. T. Ikegami and S. Sasa for stimulating discussions.
This work is partially supported by Grant-in-Aids for Scientific Research from 
the Ministry of Education,
Science and Culture of Japan.
One of authors (NK) is supported by a research fellowship from Japan Society for 
the Promotion of Science.\\

\clearpage

\begin{figure}
%\begin{center}
\noindent
\epsfig{file=m.1.eps,height=6cm,width=6cm}
\epsfig{file=m.2.eps,height=6cm,width=6cm}

\end{figure}
\clearpage

\begin{figure}
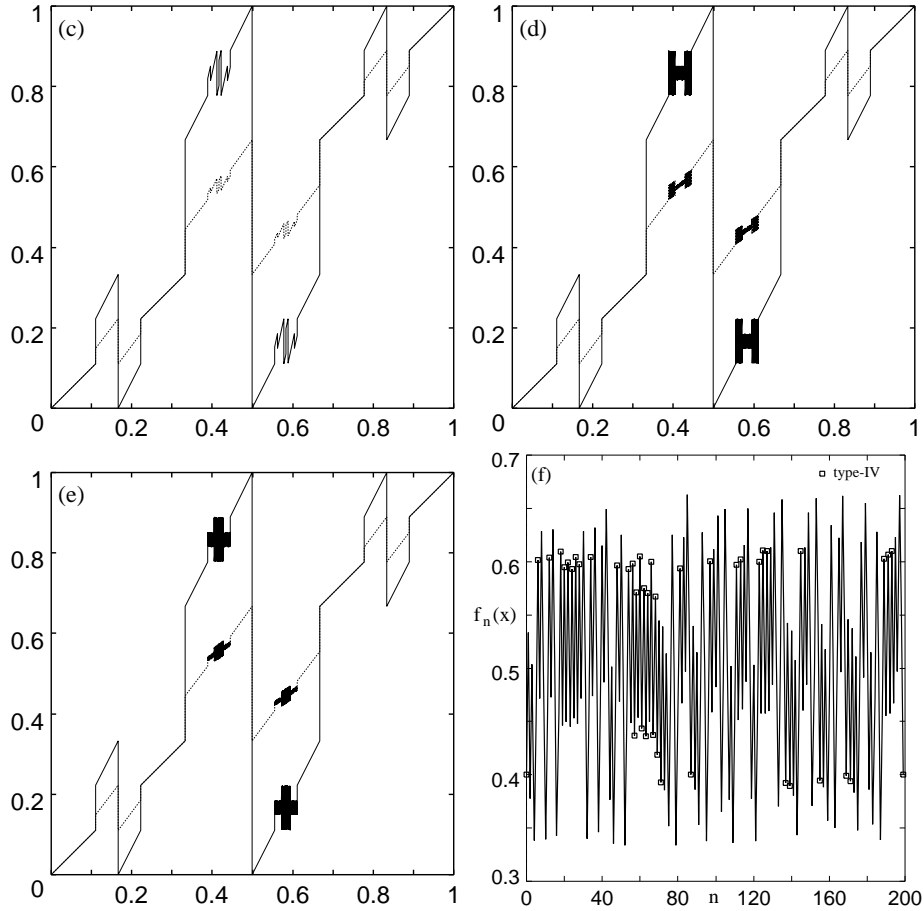

\noindent
\epsfig{file=m.3.eps,height=6cm,width=6cm}
\epsfig{file=m.10.eps,height=6cm,width=6cm}\\
\epsfig{file=m.11.eps,height=6cm,width=6cm}
\epsfig{file=mm.0-200.eps,height=6cm,width=6cm}
\caption{Time evolution of $f_n(x)$ and $g_n(x)$.
(a) $f_1(x)$ (solid line) and $g_1(x)$ (dotted line).
(b) $n = 2$.
(c) $n = 3$
(d) $n = 10$
(e) $n = 11$
(f) Time evolution of one point $f(x')$, determined by the generated meta-map, at the center square ($0 < x < 200$).
The `type' of the function $f_n(x')$ changes between III and IV in time. 
Squares indicate that the `type' of $f_n(x')$ is IV at $n$.
Since the mesh size required for the computation is extremely large,
the plotted orbit is not precise. It is expected, however, that the statistical properties are conserved 
with this numerical computation.
}
\label{fig:meta-map}
%\end{center}
\end{figure}

\begin{figure}
%\begin{center}
\noindent
\epsfig{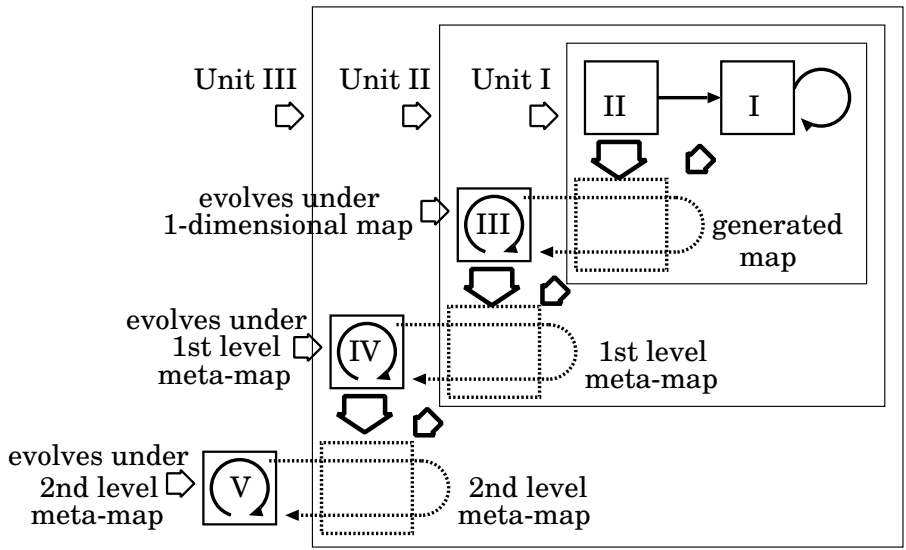}

\caption{The schema of the meta-map.
A 1-dimensional map can be generated from
the configuration of type-I and II fixed points (unit-I).
The function \(f_n(x)\) iterated by this map is a type-III point
which generates a meta-map (unit-II).
The meta-map is determined by the map generated from a type-II fixed function and determines a dynamics of a type-IV point.
The type-IV points and unit-II generate a higher level meta-map.
This process can be continued recursively.
}
\label{fig:sc2}
%\end{center}
\end{figure}

\begin{figure}
%\begin{center}
\noindent
\epsfig{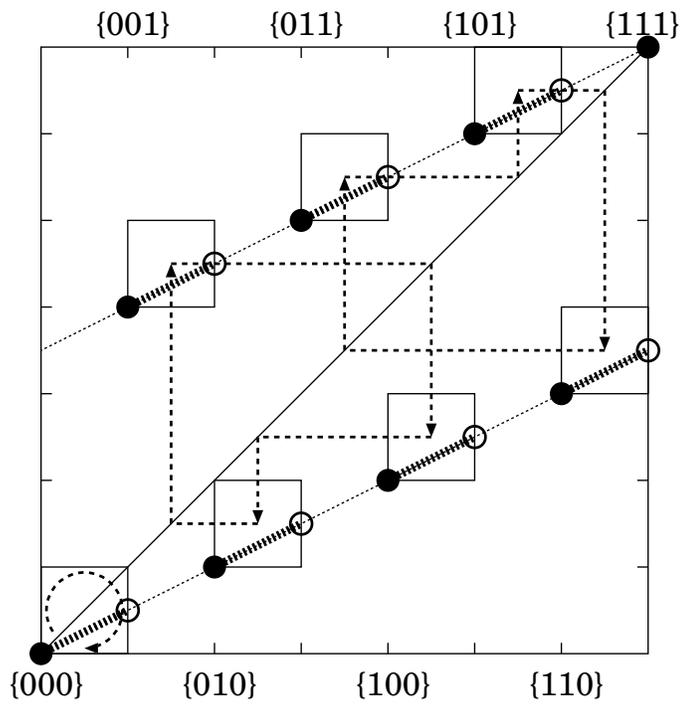}

\caption{A multi-branch Nagumo-Sato map $M_3(x)$.
Two period-3 attractors coexist.
}
\label{fig:multi2}
%\end{center}
\end{figure}

\begin{figure}
%\begin{center}
%\noindent
\epsfig{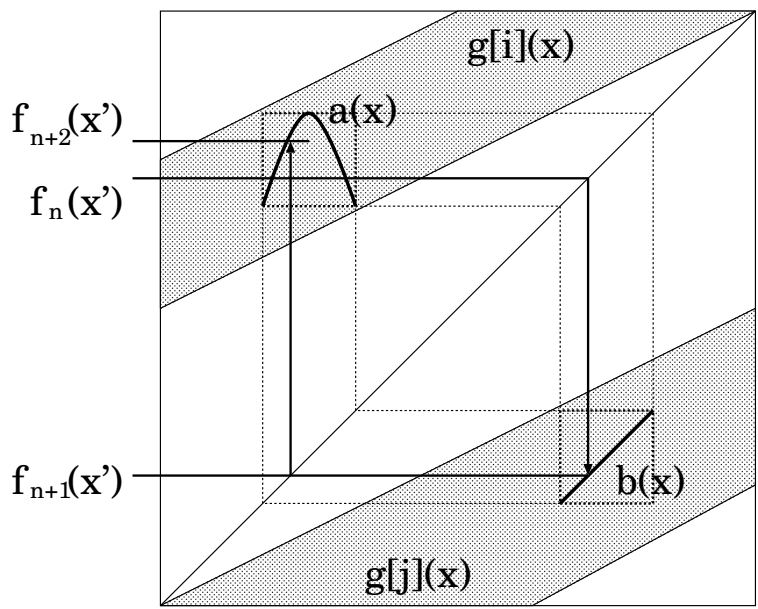}

\caption{An example with an arbitrary 1-dimensional map embedded.
Once the map is generated as in this figure,
\(f_n(x^{\prime})\), which takes a value in this region, evolves according to
\(f_{2n+1}(x^{\prime}) = a(f_{2n}(x^{\prime}))\) and
\(f_{2n+2}(x^{\prime}) = b(f_{2n+1}(x^{\prime}))
= b \circ a (f_{2n}(x^{\prime}))\).
If \(b(x)\) is the identity function,
\(f_{2n+2}(x^{\prime}) = a(f_{2n}(x^{\prime}))\).
An arbitrary 1-dimensional map
can be embedded 
by observing the dynamics of the function every two steps.}
\label{fig:logi}
%\end{center}
\end{figure}

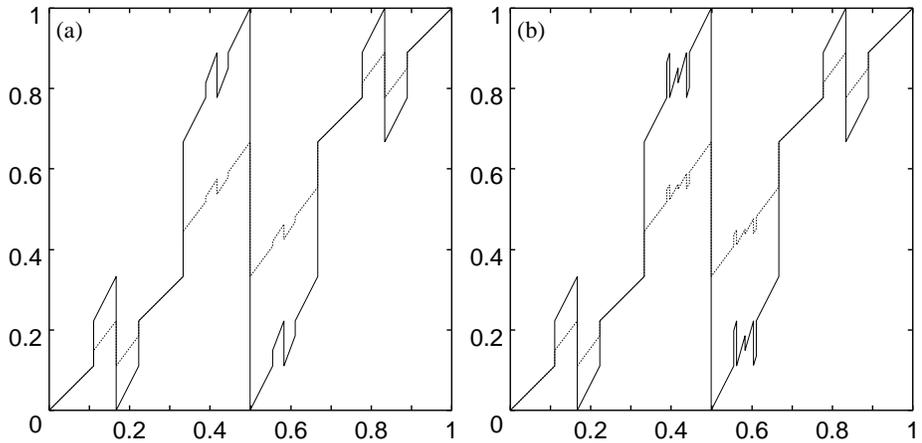
\end{document}